\documentclass[twocolumn,floatfix,aps,prl,showpacs]{revtex4}

\usepackage{SIunits}
\usepackage{multirow}
\usepackage{graphicx}
\usepackage{tabularx}
\usepackage{booktabs}
\usepackage{epsfig,graphicx}
\usepackage{flafter}
\usepackage{amsmath}
\usepackage{color}
\usepackage{braket}

\usepackage[colorlinks,urlcolor=blue,citecolor=blue,linkcolor=blue]{hyperref}

\begin{document}

\title{Collective Modes in a Unitary Fermi Gas across the Superfluid Phase Transition}

\author{Meng Khoon Tey}
\author{Leonid A. Sidorenkov}
\author{Edmundo R. S\'anchez Guajardo}
\author{Rudolf Grimm}
\affiliation{Institut f\"ur Quantenoptik und Quanteninformation (IQOQI),
 \"Osterreichische Akademie der Wissenschaften}
 \affiliation{Institut f\"ur Experimentalphysik, Universit\"at Innsbruck, 6020 Innsbruck, Austria}

\author{Mark J. H. Ku}
\author{Martin W. Zwierlein}
\affiliation{MIT-Harvard Center for Ultracold Atoms, Research Laboratory of Electronics, and Department of Physics,
Massachusetts Institute of Technology, Cambridge, Massachusetts 02139, USA}

\author{Yan-Hua Hou$^{1}$}
\author{Lev Pitaevskii$^{1,2}$}
\author{Sandro Stringari$^{1}$}
\affiliation{$^1$Dipartimento di Fisica, Universit\'a di Trento and INO-CNR BEC Center, I-38123 Povo, Italy}
 \affiliation{$^2$Kapitza Institute for Physical Problems RAS, Kosygina 2, 119334 Moscow, Russia}

\date{\today}

\begin{abstract}
We provide a joint theoretical and experimental investigation of the temperature dependence of the collective oscillations of first sound nature exhibited by a highly elongated harmonically trapped Fermi gas at unitarity, including the region below the critical temperature for superfluidity. Differently from the lowest axial breathing mode, the  hydrodynamic frequencies of the higher-nodal excitations show a temperature dependence, which is calculated starting from Landau two-fluid theory and using the available experimental knowledge of the equation of state. The experimental results agree with high accuracy with the predictions of theory and provide the first evidence for the temperature dependence  of the collective frequencies near the superfluid phase transition.

\end{abstract}
\date{\today}

\maketitle

Collective oscillations provide powerful tools to understand the physical behavior of quantum many-body systems from different points of view and to test fundamental theories. On one hand, collective modes can be used to explore different dynamical regimes of the system, such as superfluid, collisional, or collisionless regimes, for both Bose and Fermi statistics \cite{Pitaevskii2003bec, Varenna2008ufg, Giorgini2008tou, Bloch2008mbp}. On the other hand, the mode frequencies allow us to probe the equation of state (EOS) of the system, including its temperature dependence. Major benefits result from the high accuracies attainable in the measurements of collective frequencies, which often enable refined investigations of subtle interaction effects.

The many-body physics of unitary Fermi gases, i.e.\ two-component Fermi gases with infinite scattering length, has attracted tremendous interest over the past decade \cite{Varenna2008ufg, Giorgini2008tou, Bloch2008mbp}. The unitary Fermi gas is characterized by strong interaction effects in the EOS \cite{Kinast2005hco, Horikoshi2010mou, Nascimbene2010ett, Ku2012rts} and reveals a unique universal thermodynamic behavior \cite{Ho2004uto}. Furthermore, at finite temperature, the strong interactions favor the collisional hydrodynamic regime,  differently from the common situation in weakly interacting Bose gases. The low-frequency modes of a harmonically trapped Fermi gas have been the subject of intensive experimental \cite{Kinast2004efs, Bartenstein2004ceo, Altmeyer2007pmo} and theoretical (see \cite{Giorgini2008tou} and references therein) efforts. The temperature dependence has been studied in Refs.~\cite{Kinast2005doa, Wright2007ftc, Riedl2008coo}. Remarkably, at unitarity all modes observed so far turned out to be insensitive to the different nature of a superfluid and a classical gas, with their frequencies remaining independent of temperature throughout the hydrodynamic regime. Previous experiments have demonstrated the crossover from hydrodynamic to collisionless behavior, which typically occurs for temperatures approaching the Fermi temperature, without giving any further information on the regime of lower temperatures where the gas is deeply hydrodynamic and the superfluid phase transition occurs.

In this Letter, we report a joint effort of theory and experiment on higher-nodal collective modes in the unitary Fermi gas. We present a 1D hydrodynamic approach to describe axial modes in a trapped `cigar-shaped' cloud. Our experimental results confirm the predicted intrinsic sensitivity of higher-nodal modes to the EOS in the low-temperature regime, above and even well below the superfluid phase transition.

The macroscopic dynamic behavior of a superfluid is governed by the Landau two-fluid hydrodynamic equations \cite{Khalatnikov1965book} holding in the deep collisional regime $\omega\tau \ll 1$ where $\tau$ is a typical collisional time and $\omega$  is the frequency of the relevant sound mode in the trap. Landau's equations consist of  the  equation of continuity for the total density, the equation for the velocity of the superfluid component, the equation for the entropy density and the equation for the   current density. The physical  ingredients entering  these equations are the equation of state and the  superfluid density. At zero temperature they reduce to the irrotational hydrodynamic equations of superfluids, above the critical point to the usual hydrodynamic equations of normal fluids. Below $T_c$ these equations describe the propagation of first and second sound, the former being basically a density wave, with the normal and superfluid components moving in phase, the latter being a temperature or entropy wave. For weakly compressible fluids the coupling between first and second
sound is small \cite{Khalatnikov1965book}. This is the case of superfluid helium and also of the unitary Fermi gas \cite{Taylor2009fas}. Since in the present work we  investigate the collective oscillations of density (first sound), we simplify the search for the solutions of Landau's equation by requiring   that the velocity fields of the normal and superfluid components are equal. Under this approximation  the equations of motion involve only the EOS, the superfluid density playing a role  only in the propagation of second sound.

In the presence of a trapping potential, the solution of Landau's equations is highly nontrivial, due to the inhomogeneity of the density profile, and  thus far has been calculated  only for the simplest case of isotropic trapping \cite{He2007fas, Taylor2009fas}. Since the experimental excitation and observation of these  modes are more easily accessible with very elongated traps, in the following we focus on such configurations. In \cite{Bertaina2010fas} it was shown that under suitable conditions of radial trapping one can derive simplified 1D hydrodynamic equations starting from the more general 3D Landau's equations. The basic point for such a derivation is the assumption that both the velocity field $v_z$ along the long axis and the temperature fluctuations during the propagation of sound do not depend on the radial coordinates. This 1D-like hydrodynamic formulation is justified under the condition that the viscosity and the thermal conductivity are sufficiently large to ensure, respectively, the absence of radial gradients in the velocity $v_z$ and the temperature. The condition for the viscosity can be recast in the form $\eta \gg \rho_{n1} \omega$  where $\rho_{n1}$ is the normal 1D mass density, obtained by radial integration of the 3D normal density. An analogous condition holds for the thermal conductivity.  These conditions are better satisfied in the presence of tight radial confinement and for the lowest frequency oscillations.
One can estimate the values of shear viscosity $\eta$ using the experimental data of \cite{Cao2011sfp}.
For the actual conditions of our experiments both sides of the inequality are of the same order of magnitude and consequently the full applicability of the 1D hydrodynamic formulation can be justified only {\it a posteriori}. Violation of the 1D condition would result in a damping of the collective oscillations so that, to the extent that the observed damping is small, we expect that the assumption of velocity field and temperature being independent of the radial coordinate is a reasonable ansatz for our variational approach.

Under the above assumptions and focusing on the unitary Fermi gas, the equation for the velocity field, characterizing the density oscillations of the gas in a highly elongated harmonic trap, takes the form (see {\cite{hou} for a complete and systematic derivation)
 \begin{eqnarray}\label{HDT}
m(\omega^{2}-\omega^2_z)v_z-\frac{7}{5}m\omega^2_zz\partial_zv_z +\frac{7}{5}\frac{P_1}{n_1}\partial_z^2v_z =0
\end{eqnarray}
where we have considered oscillations in time  proportional to $e^{-i\omega t}$. Here, $\omega_z$ represents the trap frequency along the axial direction  $z$, and $m$ is the atom's mass.
Equation (\ref{HDT})  explicitly points out the crucial role played by the equation of state  $P_1(n_1,T)$, where $ P_1 = \int  P dxdy$ is the ``1D pressure'' (having units of force) and $n_1 = \int  n dxdy$ is the atom number per unit length. In order to derive Eq.(\ref{HDT}) we have explicitly used the adiabatic result  $n_1\left(\partial P_1/\partial n_1\right)_{\bar{s}_1}=7/5P_1$ holding at unitarity at all temperatures, where $\bar{s}_{1} = (1/n_{1})  \int s dxdy$ is the entropy per atom with $s$ the entropy density 
and we have assumed the validity of the local density approximation along both the axial and radial directions.

At zero temperature and in the classical limit of high temperature the hydrodynamic equation (\ref{HDT}) admits analytic solutions of polynomial form $v_z = a_kz^k +a_{k-2}z^{k-2}+...$, with integer values of $k$. At $T=0$,  where $P_1(n_1)/n_1 =(2/7)  [\mu_0 -(1/2)m\omega^2_zz^2]$, with $\mu_0$  the chemical potential at the center of the trap,  the frequency of the $k$-th mode is given by
\begin{equation}
\omega^2=\frac{1}{5}(k+1)(k+5)\omega_z^2 \, .
\label{HD3}
\end{equation}
In the classical limit, where $P_1/n_1=k_{B}T$, one finds the different $k$ dependence
\begin{equation}
\omega^2=\frac{1}{5}(7k+5)\omega_z^2 \, .
\label{HD4}
\end{equation}
It is worth noting that Eqs.~(\ref{HD3}) and (\ref{HD4}) coincide for $k=0$ (center of mass oscillation) and $k=1$ (lowest axial breathing mode), while they predict different values for the higher nodal solutions. One can actually prove that not only the frequency of the center of mass but, at unitarity, also the frequency of the lowest axial breathing mode are independent of temperature, corresponding to an exact scaling solution of the two fluid hydrodynamic equations \cite{hou}. 
It then follows that only the $k\ge 2$ modes exhibit a temperature dependence.

In order to provide a simple quantitative prediction for the temperature dependence of the mode frequencies,  we have developed a variational approach to the  solution of the hydrodynamic equation (\ref{HDT}) with  the ansatz $v_z= a_2 z^2 + a_0$ for the $k=2$ mode. This ansatz reproduces exactly the frequency of the $k=2$ mode in the limits of $T=0$ and high $T$. For intermediate temperatures we obtain \cite{hou}
\begin{equation}
\omega^2_{k=2}= \frac{129t_2-25}{5(9t_2-5)}\omega^{2}_{z} \, ,
\label{HDk=2}
\end{equation}
where 
$t_2= M_0M_4/M_2^2$.
We have introduced the dimensionless moments
 \begin{equation}
M_\ell= \int_{-\infty}^{\beta\mu_0}dx(\beta\mu_0-x)^{(\ell+1)/2}n(x)\lambda_{T}^{3}
\label{Ml}
\end{equation}
of the 3D number density $n(x)$, where $x$ is the local chemical potential times $\beta = 1 /k_B T$ and $\lambda_T$ is the thermal de Broglie wavelength.
The temperature dependence of the moments $M_\ell$ can be evaluated using the experimentally determined EOS \cite{Ku2012rts}. Approaching the classical regime, for $\beta\mu \lesssim -1.5$, the virial expansion of the EOS \cite{Liu2009vef} holds \cite{Ku2012rts} and allows us to extend the integral to $\beta\mu \rightarrow -\infty$. At low temperatures, corresponding to $\beta\mu > 4$, the EOS is governed by phonons and is solely determined by the Bertsch paramter $\xi$. 
The error of the quantity $t_2$ above resulting from the error in the density EOS is less than 1\%.
We have checked that our variational predictions
are practically indistinguishable from the exact numerical solution
of Eq.~(\ref{HDT}). Starting from  the equation of continuity one can also calculate the shape of the density oscillations of each mode, proportional to $\partial_z(n_1 v_z)$.

Experimentally, we prepare an ultracold, resonantly interacting Fermi gas by evaporating a two-component spin mixture of fermionic $^6$Li in an optical dipole trap \cite{Jochim2003bec}. The atomic cloud contains typically $N/2=1.5\times 10^5$ atoms per spin state, and the magnetic field is set to 834\,G, right on top of the well-known broad Feshbach resonance \cite{Chin2010fri}. For the lowest temperatures, the trapping beam (wavelength 1075 nm) has a waist of 31~$\micro$m, the trap depth is about $2\,\mu$K, and the axial and radial trap frequencies are $\omega_z=2\pi\times 22.52(2)\,$Hz and $\omega_r=2\pi\times 473(2)\,$Hz, respectively. 
For experiments at higher temperatures, the beam waist is increased to 38~$\micro$m, and deeper traps are used (up to $16\,\mu$K depth) with trap frequencies of up to $\omega_z=2\pi\times 23.31(3)\,$Hz and $\omega_r=2\pi\times 1226(6)\,$Hz. The corresponding Fermi temperatures $T_F=\hbar(3N\omega_r^2\omega_z)^{1/3}/k_B$ vary between  about $0.8$ and $1.5\,\micro$K. We point out that essentially perfect harmonic confinement along the long trap axis ($z$-axis) is ensured by the magnetic trapping that results from the curvature of the magnetic field used for Feshbach tuning~\cite{Jochim2003bec}. Also anharmonicities in the radial confinement remain negligibly small. To probe the ultracold gas we record one-dimensional axial density profiles $n_1(z)$ by near in situ absorption imaging \cite{longpaper}.

The temperature $T$ of the gas is set by controlled heating, always starting from a deeply cooled cloud ($T/T_F \approx 0.1$). In the low-temperature range ($T \lesssim 0.2\,T_F$), we simply introduce a variable hold time of up to 4\,s in which residual trap heating slowly increases $T$.
Higher temperatures are reached by parametric heating, modulating the trap power at about $2 \omega_r$, and introducing a sufficient hold time to reach thermal equilibrium between the different degrees of freedom. We characterize the resulting temperature in a model-independent way that does not require any {\it a priori} knowledge of the EOS. Based on the virial theorem \cite{Thomas2005vta} we introduce the dimensionless parameter $E/E_0$, which represents the total energy $E = 3 m \omega^2_z \int_{-\infty}^{\infty} dz z^2 n_1(z)$ 
normalized to the energy of a noninteracting, zero-temperature Fermi gas, $E_0 = \frac{3}{4} N k_B T_F$.
For a given EOS, the energy scale ($E/E_0$) can be converted to a temperature scale ($T/T_F$).
Alternatively, we obtain the cloud's temperature by fitting the experimental profiles $n_1(z)$ \cite{Ho2010otp,Nascimbene2010ett} with $T$-dependent profiles  for a given EOS. For both methods, we use the EOS from Ref.~\cite{Ku2012rts}. We note that the temperatures obtained by both methods in general show satisfying agreement with each other. At very low temperatures the latter method shows a trend to give slightly lower values of $T$ (up to $\sim$10\%), which indicates small systematic uncertainties of our measurements.

\begin{figure}
\includegraphics[width=0.85\columnwidth]{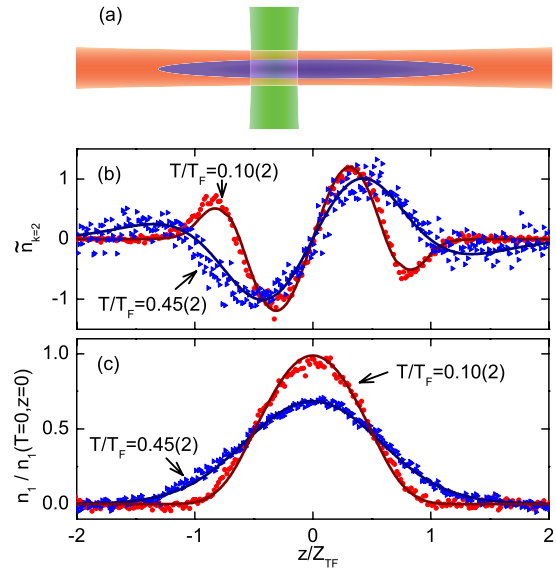}
\caption{\label{fig:profiles} (Color online) Probing a higher-order first sound longitudinal
mode at the example of $k=2$. In (a), we illustrate the basic geometry of exciting the optically trapped cloud with a weak, power-modulated repulsive laser beam, which perpendicularly intersects the trapping beam. In (b), the experimental mode profiles (data points) are compared to theoretical curves based on the experimental EOS from \cite{Ku2012rts} (solid lines) for two different temperatures. The corresponding cloud profiles in (c) are analyzed to extract the temperatures (see text). The solid lines show the density profiles obtained from the EOS \cite{Ku2012rts} with $T/T_F = 0.10$ and $0.45$. The parameter $Z_{\rm TF} = \xi^{1/4} \omega^{-1}_z \sqrt{2 k_B T_F / m}$ represents the Thomas-Fermi radius of the zero-$T$ interacting gas.}

\end{figure}

We selectively excite axial modes of order $k$ by using a resonant excitation scheme. As illustrated in Fig.~\ref{fig:profiles}(a), a repulsive 532-nm laser beam perpendicularly intersects the trapping beam, with its position and size chosen in a way to provide best mode matching. Typically, the excitation pulse contains $8$ cycles of sinusoidal modulation with a half-cycle sine envelope, and the maximum potential height of the excitation beam is kept to about $0.01\,k_B T_F$. The power, length and shape of the excitation pulse are optimized in order to resonantly drive the desired small-amplitude oscillation. The amplitude of the corresponding density modulation stays well below 3\% of the central density of the cloud.

We record axial density profiles $n_1(z,t)$ of the excited cloud for various time delays $t$ after the excitation pulse. We then perform a Fourier transform (FFT). The resulting function $\tilde{n}(z,\omega)$ reveals the collective mode spectrum with eigenfrequencies $\omega_k$ and the corresponding spatial mode profiles $\tilde{n}_k(z)$. We extract the precise frequency of a particular mode $k$ by projecting $n_1(z,t)$ onto the spatial profile $\tilde{n}_k(z)$ and analyzing the resulting oscillation in the time domain \cite{longpaper}. The high signal-to-noise ratio results in very low statistical uncertainties in the permille range.

In this way, we study the longitudinal modes with $k=0, 1, 2$. The measured frequency $\omega_0$ of the sloshing mode ($k=0$) is an accurate measure of the axial trap frequency ($\omega_{k=0} = \omega_z$), and is therefore used for normalization purposes. The axial compression mode ($k=1$) has been studied in previous work \cite{Bartenstein2004ceo, Nascimbene2009coo}. Here, in the full temperature range explored ($0.1\le T/T_F \le 0.5$) we observe its frequency very close to $\omega_{k=1} = \sqrt{12/5} \, \omega_z$. Deviations from this value remain below 0.3\% and no significant temperature dependence is observed. This confirms that this mode is insensitive to the temperature as long as the gas stays hydrodynamic. For the higher-nodal mode with $k=2$, we observe the expected $T$-dependent frequency variations. Damping increases as compared to the $k=1$ mode, but remains sufficiently small to observe many oscillations and thus to accurately determine the mode frequency. Typical observed damping times \cite{longpaper} are 2\,s at 0.5\,$T_c$, 0.2\,s at $T_c$, and 0.12\,s at 2\,$T_c$. We note that the $k=3$ mode \cite{longpaper} shows very similar behavior, with larger frequency variations but faster damping.


In Figs.~\ref{fig:profiles}(b) and (c), we show examples for the spatial profiles of the $k=2$ mode for two different temperatures $T/T_F = 0.10$ and $0.45$ along with the corresponding unperturbed density profiles of the cloud. The comparison of the experimental data (data points) with the theoretical results based on the experimental EOS in \cite{Ku2012rts} (solid lines) shows excellent agreement.

\begin{figure}
\includegraphics[width=1.0\columnwidth]{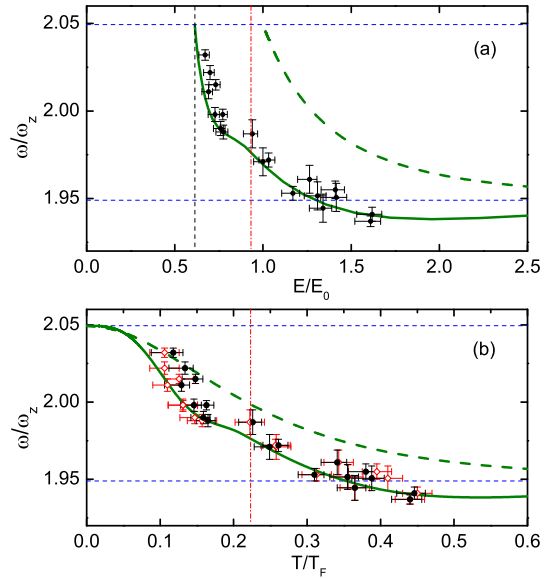}
\caption{\label{fig:modefreq} (Color online) Comparison between experimental and theoretical first sound frequencies of the $k=2$ mode. In (a) the experimental data are plotted versus the energy parameter $E/E_0$, while in (b) we use a temperature scale $T/T_F$. The theoretical curves (solid lines) are based on the EOS of Ref.~\cite{Ku2012rts}. This EOS is also used to extract $T/T_F$ from the measured profile in two different ways: The filled symbols in (b) result from a direct conversion of $E/E_0$ to $T/T_F$, while the open symbols result from fitting the cloud profiles (see text). For comparison, we also show the mode frequencies (dashed curves) that would result from the EOS of the ideal Fermi gas. The thin horizontal dashed lines mark the zero-$T$ superfluid limit ($\omega/\omega_z = \sqrt{21/5}$) and the classical hydrodynamic limit ($\omega/\omega_z = \sqrt{19/5}$) according to Eqs.~(\ref{HD3}) and (\ref{HD4}), respectively. In (a) the dashed vertical line indicates the $T=0$ ground state with $E/E_0 = \sqrt{\xi} = 0.613(3)$, while the dash-dotted vertical lines in (a) and (b) indicate the critical energy $E_c/E_0 = 0.934(39)$ and temperature $T_c/T_F = 0.223(15)$.}
\end{figure}

Figure~\ref{fig:modefreq} presents the comparison between the experimental and theoretical frequencies for the $k=2$ mode. In (a) the normalized mode frequencies $\omega_{k=2}/\omega_z$ are plotted versus the energy parameter $E/E_0$, while (b) displays the same data on a temperature scale $T/T_F$. The experimental data confirm the pronounced temperature dependence of the mode frequencies as predicted by our theory based on the EOS of Ref.~\cite{Ku2012rts} (solid line). In comparison, the disagreement with the dependence that would result from the EOS of the ideal Fermi gas (dashed line) highlights the important role of the EOS. At the lowest temperatures ($T/T_F \approx 0.1$) the frequency lies close to the $T=0$ superfluid limit, but already shows a significant down-shift amounting to almost 1\%. At the highest temperatures ($T/T_F \approx 0.45$) our data show a clear trend to go below the asymptotic high-temperature value, i.e.\ the classical hydrodynamic limit. The corresponding nonmonotonic temperature dependence can be understood based on the first-order correction to the EOS resulting from the virial expansion at high temperatures.

In conclusion, our combined theoretical and experimental work on higher-nodal axial collective modes of a unitary Fermi gas reveals a pronounced temperature dependence below and near the superfluid phase transition. The observed temperature dependence is a unique feature of higher-nodal modes, not present for any other collective mode studied in Fermi gases so far. Our theoretical approach is based on a 1D two-fluid hydrodynamic model describing the frequently used elongated `cigar-shaped' trap geometry. The excellent agreement with the experimental results provides a stringent test for the validity of this 1D approach and highlights its potential power to accurately describe second sound modes. Moreover, our measurements provide an independent confirmation of the recently measured EOS of a unitary Fermi gas.

We would like to thank John Thomas for useful discussions and Florian Schreck for discussion and experimental support. The Innsbruck team acknowledges support from the Austrian Science Fund (FWF) within SFB FoQuS (Project No.\ F4004-N16). The Trento team acknowledges support from the European Research Council through the project QGBE. The MIT work was supported by the NSF, AFOSR, ONR, ARO with funding from the DARPA OLE program, and the David and Lucile Packard Foundation.


\end{document}